\documentclass[a4paper,11pt]{article}
\usepackage[dvips]{graphicx}
\usepackage{amssymb}
\usepackage{amsmath}

\usepackage{cite}

\begin{document}

\title{One Dimensional Supersymmetric Algebras in Color Superconductors and Reissner-Nordstr\"{o}m-anti-de Sitter Gravitational Systems}
\author{V. K. Oikonomou\thanks{
Vasilis.Oikonomou@mis.mpg.de}\\
Max Planck Institute for Mathematics in the Sciences\\
Inselstrasse 22, 04103 Leipzig, Germany} \maketitle

\begin{abstract}
We study two fermionic systems that have an underlying
supersymmetric structure, namely a color superconductor and Dirac
fermion in a Reissner-Nordstr\"{o}m-anti-de Sitter gravitational
background. In the chiral limit of the color superconductor, the
localized fermionic zero modes around the vortex form an $N=2$ with
zero central charge $d=1$ quantum algebra, with all the operators being Fredholm. We compute the Witten index of this algebra and we find an unbroken supersymmetry.
The fermionic gravitational
system in the chiral limit too, has two underlying unbroken $N=2$, $d=1$ supersymmetric
algebras. The unbroken supersymmetry in the later is guaranteed by
the existence of fermionic quasinormal modes in the
Reissner-Nordstr\"{o}m-anti-de Sitter background. In this case the operators are not Fredholm and regularized indices are deployed.
\end{abstract}

\section*{Introduction}

During the last decade the research towards the interrelation of
gravity and condensed matter physical
 systems has received considerable attention, especially the study of holography in such systems \cite{holography1,holography2,holography3,holography4,holography5,holography6}.
 This research stream was further enhanced by the experimental verification of certain condensed matter systems that have topological
 origin \cite{nishida}, an observation which was done very recently actually \cite{experiments1,experiments2,experiments3}.
 Particularly these where time reversal symmetric extensions of the famous topological originating quantum Hall effect in two \cite{ti2,ti21} (topological quantum spin Hall effect) and three dimensions \cite{ti3,ti31,ti32}, known as topological insulators \cite{nishida}. Of course, the quantum Hall effect is the most famous topologically non trivial state of matter in two dimensions \cite{nightingale}. A non-trivial charge of the single-particle Hamiltonian is intrinsic to a topological state of matter. Along with the topological insulators, the topological superconductors serve as another class of topological states of condensed matter, particularly the $p_x+ip_y$ ones. Moreover gapless localized fermions appear around the vortex core of a vortex defect in topological superconductors. Such theoretical constructions are realized in the surface of three dimensional insulators \cite{nishida,fu}. The most extreme condensed matter states can occur at ultra high quark densities. Of course this is no ordinary 
matter since the quarks are deconfined, but we are talking about condensed matter physics of QCD. At ultra high densities the QCD coupling is relatively small, a situation that can physically occur in neutron stars for example. In these conditions (low temperature, high density) the quarks may form cooper-like pairs, thus breaking explicitly the gauge symmetry, and forming a so-called color superconductor. Bearing in mind that matter states can be topological in ordinary superconductors, it is natural to ask whether a
 color superconductor is topological. This questions has been answered and the answer is in the affirmative \cite{nishida}. In
 relation to this cooper pair condensation, a color superconductor might be related to a condensation that fermionic fields might
experience in involved AdS gravitational backgrounds. A similar phenomenon to
a holographic superconducting phase transition
\cite{holography1,holography2,holography3,holography4,holography5,holography6} was studied in
\cite{cai}, but for an Reissner-Nordstr\"{o}m-anti-de Sitter black
hole spacetime, and the results where consistent with the
holographic superconducting phase transition. Particularly in
\cite{cai} the order parameter is an Dirac fermion charged through
a direct coupling to a Maxwell field. In this sense, this model of
a charged Dirac fermion in the background of an
Reissner-Nordstr\"{o}m-anti-de Sitter black hole, could serve as a
simple model of color superconductivity. The approach of the
authors of \cite{cai} is done by computing the quasinormal modes
of the charged Dirac fermion field in the aforementioned curved
background. As is well known, quasinormal modes
\cite{kokkotas,review,reviewqm1,roman}, describe a long lasting
period of damped  gravitational waves oscillations. The
quasinormal modes are so to speak, the characteristic sound of
black holes hence matter field perturbations of such gravitational
backgrounds can be a useful tool to study black holes. Moreover
the quasinormal modes depend only on a few physical parameters of
the black hole, namely, the mass, angular momentum and charge of
the black holes, thus rendering the spacetime parameters
identification easier. The perturbation of a quantum field in a
black hole background consists of three time evolution stages,
that is, the wave burst, the quasinormal mode stage and the
power-law tail \cite{quasidirac5}. For the calculation of a Dirac fermion quasinormal modes in a Reissner-Nordstr\"{o}m-anti-de Sitter Spacetimes in $D=4$ and $D>4$ we refer the reader to references \cite{quasidirac5a}, and \cite{quasidirac5b} respectively. In addition for Dirac quasinormal modes in curved backgrounds see also \cite{quasidirac5c,quasidirac5d,quasidirac5e}. Owing to the vast number of
applications and implications of many theoretical frameworks that
embody quasinormal modes, research in this area has attracted a
lot of attention. Firstly the existence of a black hole
 can be directly verified by observing it's fundamental quasinormal mode.
Additionally the thermodynamic properties of loop quantum gravity
(an appealing alternative to string theory) black holes can be
further understood using the quasinormal modes. Moreover the quasinormal modes of anti de
Sitter black holes have a dual physical correspondence to
quantities of the dual conformal field theory via the well known
AdS/CFT correspondence \cite{kanti}. From astrophysical aspects,
the most interesting spacetimes are the asymptotically flat ones,
however the observation of the universe's expansion motivated the
study of quasinormal modes in de Sitter \cite{quasidirac7}. Quasinormal modes can yield which
gravitational systems are stable under dynamical perturbations.
Actually a static or non-static solution describing a compact
object is stable if all it's quasinormal modes are decaying in
time, on the contrary even if one mode is growing, the
gravitational system is unstable \cite{roman}.

Owing to the fact that the charged Dirac fermion in the background
of an Reissner-Nordstr\"{o}m-anti-de Sitter black hole could be a
simple model of color superconductivity, we present in this paper
that in both systems, namely the fermionic spectrum around the
boundary vortex of a color superconductor and the fermion in the
Reissner-Nordstr\"{o}m-anti-de Sitter black hole spacetime, there
exists a hidden $N=2$ supersymmetric quantum mechanics algebra
\cite{witten,thaller,susyqm,susyqm1} (SUSY QM hereafter).
Particularly, for the color superconductor system, the
supersymmetric algebra occurs for the $m=0$, $p_z=0$ (chiral case)
case of the Bogoliubov-de Genne equation. In the case of the
fermionic field in the curved gravitational background, the
supersymmetric algebra is very closely related to the quasinormal
modes spectrum, and the very existence of supersymmetry is
guaranteed by the existence of quasinormal modes. In the case of
color superconductivity, the supersymmetry is due to the vortex,
which actually causes localized fermionic solutions around it.
Supersymmetric structures of the same kind around defects where
studied in \cite{oikonomou1,oikonomoustrings} where the case of a
superconducting and a cosmic string where analyzed respectively.
Supersymmetry in the case where fermionic quasinormal modes are
studied in various gravitational backgrounds, was investigated in
\cite{oikonomou3}. In the case we shall present in this paper, the
fermionic system actually has two $N=2$ $d=1$ supersymmetries, the
supercharges of which could be related to harmonic superspace
extensions
\cite{ivanov1,ivanov2,ivanov3,ivanov4,ivanov5,ivanov6,ivanov7,ivanov8,ivanov9}.
The supersymmetric quantum mechanics algebras are very important
from physical and mathematical point of views, since these can be
directly connected to harmonic superspace
\cite{ivanov1,ivanov2,ivanov3,ivanov4,ivanov5,ivanov6,ivanov7,ivanov8,ivanov9}
and to $d=1$ supersymmetric sigma models with very interesting
target space geometries. Furthermore, these supersymmetric
extensions provide superextensions of the Landau problem and of
the quantum Hall effect \cite{ivanovlan,ivanovlan1,ivanovhal}. In addition $N=2$ $d=1$ supersymmetry appears in condensed matter systems, like in graphene for example, see \cite{graphene}. In addition, there is a close connection of photon, gravitino and graviton modes from extremal Reissner-Nordstr\"{o}m black holes, which is expressed in terms of an isospectrality in the spectrum \cite{re1,re2,re3}. Although these works investigate systems in the context of supergravity, a supersymmetric quantum mechanics algebra could be a remnant of local supersymmetry. It is surprisingly interesting that the color superconductors and the
fermionic system in Reissner-Nordstr\"{o}m-anti-de Sitter black
hole (a model that is believed to the gravitational description of
color superconductivity) are linked via the same supersymmetric
underlying pattern. However, these supersymmetries are different,
owing to the fact that in the case of color superconductors, the
operators are Fredholm, while in the case of the gravitational
system that is not true.

This paper is organized as follows: In section 1 we present the
color superconductor model and the underlying $N=2$ SUSY QM
algebra. In section 2 we study the  charged Dirac fermion in the
background of an Reissner-Nordstr\"{o}m-anti-de Sitter black hole
and we present the structure of the resulting two SUSY QM. At the end of section 2 we present a
global symmetry that the aforementioned fermionic system possess.
The conclusions follow thereafter.

\section{Superconductors and Vortices}

In this section we study the underlying supersymmetry that the
fermionic system that describes color superconductivity has. We
start with the mean-field model of color superconductivity, it's
benchmark of which is the Hamiltonian \cite{nishida}:
\begin{align}\label{initialhamiltonian}
&H_{CSC}=\int \mathrm{d}x \Big{[} \psi_{a,f}^{\dag}(-i\alpha
\partial +\beta m-\mu )\delta_{ab}\delta_{fg}\psi_{b,g}
\\ \notag & +\frac{1}{2}\psi^{\dag}_{a,f}\Delta_{ab,fg}(x)C\gamma^5\psi^*_{b,g}
-\frac{1}{2}\psi^{T}_{a,f}\Delta_{ab,fg}^{\dag}(x)C
\gamma^5\psi_{b,g} \Big{]}
\end{align}
with $\alpha,\beta$ and $\gamma_5$ being equal to:
\begin{equation}\label{arxchiralrep}
\alpha=\gamma^{0}\gamma^{5}=\bigg{(}\begin{array}{ccc}
  \sigma & 0 \\
  0 & -\sigma  \\
\end{array}\bigg{)},{\,}{\,}{\,}{\,}\beta=\gamma^0=\bigg{(}\begin{array}{ccc}
  0 & I \\
  I & 0  \\
\end{array}\bigg{)},{\,}{\,}{\,}{\,}\gamma^{5}=\bigg{(}\begin{array}{ccc}
  I & 0 \\
  0 & -I  \\
\end{array}\bigg{)}
\end{equation}
In the above equation (\ref{initialhamiltonian}), the matrix $C$
stands for the charge conjugation matrix, namely
$C=i\gamma^2\gamma^{0}$, where $\gamma^i$ are the Dirac gamma
matrices. The model that is described by the aforementioned
Hamiltonian, contains three colors and three flavors, which are
denoted by the letters $a,b$ and $f,g$ respectively, in the
Hamiltonian (\ref{initialhamiltonian}). The pairing gap is
described by $\Delta_{ab,fg}$, in the Lorentz singlet and even
parity channel ($J^p=0^+$). Its specific dependence on the color
and flavor is described by:
\begin{equation}\label{fug}
\Delta_{ab,fg}(x)=\sum_{i=1,2,3}\Delta_{i}\epsilon_{iab}\epsilon_{ifg}
\end{equation}
The Hamiltonian (\ref{initialhamiltonian}) after a orthogonal
transformation in the color-flavor space, can be brought in the
decoupled form, with $H_{SCS}=\sum_i^9H_j$, with $H_j$ being equal
to:
\begin{align}\label{hamiltonian}
& H_j=\int\mathrm{d}\left [ \psi_j^{\dag}(-ia\partial+\beta
m-\mu)\psi_j+\frac{1}{2}\psi^{\dag}_j\Delta(x)_jC\gamma^5\psi^*_j
-\frac{1}{2}\psi^{T}_j\Delta^*(x)_jC\gamma^5\psi_j\right ] \\
\notag & = \frac{1}{2}\int\mathrm{d}\left (
\psi_j^{\dag}-\psi_j^{T}C\gamma^5 \right )
 \left
(\begin{array}{ccc}
  -i\mathbf{a}\mathbf{\partial}+\beta m-\mu & \Delta_j(x) \\
  \Delta_j^*(x) & i\mathbf{a\partial}-\beta m+\mu  \\
\end{array}\right )
\left(%
\begin{array}{c}
  \psi_j \\
  C\gamma^5\psi_j^* \\
\end{array}%
\right)
\\ \notag & =\frac{1}{2}\int
\mathrm{d}\Psi^{\dag}_j\mathcal{H}_j\Psi_j
\end{align}
The case where $\Delta \neq 0$ describes a fully gapped color
flavor locked phase. The Hamiltonian (\ref{hamiltonian}) possesses
many symmetries, like the charge conjugation symmetry and the time
reversal symmetry. Such Hamiltonian have symmetry properties that
have been classified and tabulated formally, see for example
\cite{zirn,altland}. The single particle Hamiltonian $H$ has the
following charge conjugation symmetry:
\begin{equation}\label{chargeconj}
\mathcal{C}^{-1}H\mathcal{C}=-H^{*}
\end{equation}
with $\mathcal{C}$ being equal to:
\begin{equation}\label{cm}
\left (\begin{array}{ccc}
  0 & -C\gamma^5 \\
  C\gamma^5 &  0 \\
\end{array}\right )
\end{equation}
Moreover when $\Delta(x)$ is a real number and in addition has a
uniform phase over the space, the Hamiltonian has the following
transformation properties:
\begin{equation}\label{timereversal}
T^{-1}HT=H^{*}
\end{equation}
where $T$ stands for:
\begin{equation}\label{tm}
T=\left (\begin{array}{ccc}
 \gamma^1\gamma^3  &  0\\
 0  &  \gamma^1\gamma^3
\end{array}\right )
\end{equation}
It is a common fact in the superconductor literature that for an
$2D$ $p_x+ip_y$ superconductor, the non-trivial topological charge
of the free space Hamiltonian is closely to a localized fermionic
zero mode around a vortex line (see \cite{nishida} and references therein). Same arguments hold for the
Hamiltonian (\ref{hamiltonian}). We are interested in zero mode
fermionic solutions around vortices, with a non-trivial pairing
gap, in the even parity pairing case \footnote{For more details on
the specifics of the superconductors see references \cite{review1,alford}
and references therein.}. The theoretical context that underlies
the calculation of the fermionic spectrum around a quantized
vortex line is pretty much described by the Bogoliubov-de Genne
equation, namely:
\begin{equation}\label{beta1}
\left (\begin{array}{ccc}
  -i\mathbf{a}\mathbf{\partial}+\beta m-\mu & e^{i\theta}|\Delta(r)| \\
  e^{-i\theta}|\Delta(r)| & i\mathbf{a\partial}-\beta m+\mu  \\
\end{array}\right )\Phi(x)=E\Phi(x)
\end{equation}
where we employed polar coordinates to be our coordinate system.
The above Hamiltonian describes the free space one particle
Hamiltonian with pairing gap $\Delta(x)= e^{i\theta}|\Delta(r)|$
and under the assumption that the vortex line extents in the
$z$-direction and also that the pairing gap does not depend on
$z$. Additionally, $|\Delta(r)|$ is required to obey
$\lim_{r\rightarrow \infty} |\Delta(r)|> 0$, or in words it is
required to have a positive non-vanishing asymptotic value. Hence
any localized fermion solutions that we will find in the following, can
be considered independent of the vortex, a fact that entails some sort of
universality of the solutions (see also the comment at end of the present section). The zero modes we shall present
have a purely topological origin \cite{nishida} in contrast to
other solutions describing bound fermions of Caroli-de
Gennes-Matricon type with vortex dependent solutions
\cite{nishida,caroli}. The solution to the above equation
(\ref{beta1}) look like:
\begin{equation}\label{solutionsgeneral}
\Phi(r,\theta,z)=e^{ip_z{\,}z}\phi_{p_z} (r,\theta)
\end{equation}
We shall be mainly interested in the case $m=0$ and $p_z=0$, and
particularly in the zero energy Bogoliubov-de Genne equation at
$m=p_z=0$. The last case is the so-called chiral limit, in
reference to $m=0$. The solutions of this equation will actually
be the localized zero modes around the vortex line. We can
classify the solutions of the zero mode (E=0) Bogoliubov-de Genne
equation to left handed and right handed fermion solutions
according to their $\gamma_5$ parity. These solutions are
exponentially localized solutions around the vortex and are equal
to \cite{nishida}:
\begin{equation}\label{rsymetry}
\varphi_R=\frac{e^{i\frac{\pi}{4}}}{\sqrt{\lambda}}\left
(\begin{array}{c}
   J_0(\mu r)\\
    ie^{i\theta}J_1(\mu r) \\
    0\\
    0 \\
    e^{-i\theta}J_1(\mu r) \\
    -iJ_0(\mu r) \\
   0\\
    0 \\
\end{array}\right )e^{-\int_0^r|\Delta (r')|\mathrm{d}r'}
\end{equation}
in reference to the right handed one, while the left handed one
takes the form \cite{nishida}:
\begin{equation}\label{rsymetry1}
\varphi_L=\frac{e^{-i\frac{\pi}{4}}}{\sqrt{\lambda}}\left
(\begin{array}{c}
    0\\
    0 \\
   J_0(\mu r)\\
    -ie^{i\theta}J_1(\mu r) \\
    0\\
    0 \\
    e^{-i\theta}J_1(\mu r) \\
    iJ_0(\mu r) \\

\end{array}\right )e^{-\int_0^r|\Delta (r')|\mathrm{d}r'}
\end{equation}
The above fermionic system, which is based on the zero modes
solutions of the Bogoliubov-de Genne equation, namely:
\begin{equation}\label{zerobdg}
\left (\begin{array}{ccc}
  -i\mathbf{a}\mathbf{\partial}+\beta m-\mu & e^{i\theta}|\Delta(r)| \\
  e^{-i\theta}|\Delta(r)| & i\mathbf{a\partial}-\beta m+\mu  \\
\end{array}\right )\Phi(x)=0
\end{equation}
can constitute an $N=2$ supersymmetric quantum mechanics algebra
($N=2$ SUSY QM hereafter). To see this, let us briefly present the
basic features of an unbroken $N=2$ SUSY QM algebra. The
generators of the $N=2$ algebra are the two supercharges $Q_1$ and
$Q_2$ and a Hamiltonian $H$, which obey \cite{witten,thaller,susyqm,susyqm1},
\begin{equation}\label{sxer2}
\{Q_1,Q_2\}=0,{\,}{\,}{\,}H=2Q_1^2=2Q_2^2=Q_1^2+Q_2^2
\end{equation}
The supercharges can be used to define the new supercharge,
\begin{equation}\label{s2}
{\mathcal{Q}}=\frac{1}{\sqrt{2}}(Q_{1}+iQ_{2})
\end{equation}
and the its adjoint,
\begin{equation}\label{s255}
{\mathcal{Q}}^{\dag}=\frac{1}{\sqrt{2}}(Q_{1}-iQ_{2})
\end{equation}
The new supercharges satisfy,
\begin{equation}\label{s23}
Q^{2}={Q^{\dag}}^2=0
\end{equation}
and additionally,
\begin{equation}\label{s4}
\{{\mathcal{Q}},{\mathcal{Q}}^{\dag}\}=H
\end{equation}
A very important element of the algebra is the Witten parity, $W$,
defined as,
\begin{equation}\label{s45}
[W,H]=0
\end{equation}
which anti-commutes with the supercharges,
\begin{equation}\label{s5}
\{W,{\mathcal{Q}}\}=\{W,{\mathcal{Q}}^{\dag}\}=0
\end{equation}
Additionally $W$ satisfies the following,
\begin{equation}\label{s6}
W^{2}=1
\end{equation}
The main utility of the Witten parity $W$, is that it spans the
Hilbert space $\mathcal{H}$ of the quantum system to positive and
negative Witten parity subspaces, that is,
\begin{equation}\label{shoes}
\mathcal{H}^{\pm}=P^{\pm}\mathcal{H}=\{|\psi\rangle :
W|\psi\rangle=\pm |\psi\rangle \}
\end{equation}
Hence, the quantum system Hilbert space $\mathcal{H}$ can be
written $\mathcal{H}=\mathcal{H}^+\oplus \mathcal{H}^-$. For the
present case we shall choose a specific representation for the
operators defined above, which for the general case can be
represented as:
\begin{equation}\label{s7345}
W=\bigg{(}\begin{array}{ccc}
  I & 0 \\
  0 & -I  \\
\end{array}\bigg{)}
\end{equation}
with $I$ the $N\times N$ identity matrix. Recalling that ${\mathcal{Q}}^2=0$
and $\{{\mathcal{Q}},W\}=0$, the supercharges can take the form,
\begin{equation}\label{s7}
{\mathcal{Q}}=\bigg{(}\begin{array}{ccc}
  0 & A \\
  0 & 0  \\
\end{array}\bigg{)}
\end{equation}
and
\begin{equation}\label{s8}
{\mathcal{Q}}^{\dag}=\bigg{(}\begin{array}{ccc}
  0 & 0 \\
  A^{\dag} & 0  \\
\end{array}\bigg{)}
\end{equation}
Consequently,
\begin{equation}\label{s89}
Q_1=\frac{1}{\sqrt{2}}\bigg{(}\begin{array}{ccc}
  0 & A \\
  A^{\dag} & 0  \\
\end{array}\bigg{)}
\end{equation}
and also,
\begin{equation}\label{s10}
Q_2=\frac{i}{\sqrt{2}}\bigg{(}\begin{array}{ccc}
  0 & -A \\
  A^{\dag} & 0  \\
\end{array}\bigg{)}
\end{equation}
The $N\times N$ matrices $A$ and $A^{\dag}$, serve as annihilation
and creation operators, with, $A: \mathcal{H}^-\rightarrow
\mathcal{H}^+$ and also $A^{\dag}$ as, $A^{\dag}:
\mathcal{H}^+\rightarrow \mathcal{H}^-$. Based on relations
(\ref{s7345}), (\ref{s7}), (\ref{s8}) the Hamiltonian $H$, can
take a diagonal form,
\begin{equation}\label{s11}
H=\bigg{(}\begin{array}{ccc}
  AA^{\dag} & 0 \\
  0 & A^{\dag}A  \\
\end{array}\bigg{)}
\end{equation}
Hence the total supersymmetric Hamiltonian $H$ that describes the
supersymmetric system, can be written in terms of the superpartner
Hamiltonians,
\begin{equation}\label{h1}
H_{+}=A{\,}A^{\dag},{\,}{\,}{\,}{\,}{\,}{\,}{\,}H_{-}=A^{\dag}{\,}A
\end{equation}
For reasons that will be immediately clear, we define the operator
$P^{\pm}$, the eigenstates of which, $|\psi^{\pm}\rangle$, satisfy
the following relation:
\begin{equation}\label{fd1}
P^{\pm}|\psi^{\pm}\rangle =\pm |\psi^{\pm}\rangle
\end{equation}
Therefore we call them positive and negative parity eigenstates,
parity referring to the $P^{\pm}$ operator. Representing the
Witten operator as in (\ref{s7345}), the parity eigenstates can be
cast in the following representation,
\begin{equation}\label{phi5}
|\psi^{+}\rangle =\left(%
\begin{array}{c}
  |\phi^{+}\rangle \\
  0 \\
\end{array}%
\right)
\end{equation}
and also,
\begin{equation}\label{phi6}
|\psi^{-}\rangle =\left(%
\begin{array}{c}
  0 \\
  |\phi^{-}\rangle \\
\end{array}%
\right)
\end{equation}
with $|\phi^{\pm}\rangle$ $\epsilon$ $H^{\pm}$. Using the
formalism we just exploited, we construct an $N=2$ SUSY QM algebra
using the fermionic system around the vortex. The Bogoliubov-de
Genne equation can be written as:
\begin{equation}\label{zerobdg}
\mathcal{D}\Phi(x)=0
\end{equation}
with $\mathcal{D}$ being equal to:
\begin{equation}\label{bdgeqinmform}
\mathcal{D}=\left (\begin{array}{ccc}
  -i\mathbf{a}\mathbf{\partial}-\mu & e^{i\theta}|\Delta(r)| \\
  e^{-i\theta}|\Delta(r)| & i\mathbf{a\partial}+\mu  \\
\end{array}\right )
\end{equation}
Based on the above matrix, we can built a supersymmetric algebra.
Indeed, the adjoint of $\mathcal{D}$ is equal to,
\begin{equation}\label{bdgeqinmformadj}
\mathcal{D}^{\dag}=\left (\begin{array}{ccc}
  i\mathbf{a}\mathbf{\partial}-\mu & e^{i\theta}|\Delta(r)| \\
  e^{-i\theta}|\Delta(r)| & -i\mathbf{a\partial}+\mu  \\
\end{array}\right )
\end{equation}
The zero modes equation for this matrix is $\mathcal{D}^{\dag}\Phi'(x)=0$.
The supercharges of the SUSY QM algebra, ${\mathcal{Q}}$ and ${\mathcal{Q}}^{\dag}$ can be
defined in terms of  $\mathcal{D}$ and $\mathcal{D}^{\dag}$ as follows,
\begin{equation}\label{wit2}
{\mathcal{Q}}=\bigg{(}\begin{array}{ccc}
  0 & \mathcal{D} \\
  0 & 0  \\
\end{array}\bigg{)}, {\,}{\,}{\,}{\,}{\,}{\mathcal{Q}}^{\dag}=\bigg{(}\begin{array}{ccc}
  0 & 0 \\
  \mathcal{D}^{\dag} & 0  \\
\end{array}\bigg{)}
\end{equation}
Moreover, the quantum Hamiltonian of the SUSY QM system is,
\begin{equation}\label{wit4354}
H=\bigg{(}\begin{array}{ccc}
  \mathcal{D}\mathcal{D}^{\dag} & 0 \\
  0 & \mathcal{D}^{\dag}\mathcal{D}  \\
\end{array}\bigg{)}
\end{equation}
It is easy to see that the supercharges (\ref{wit2}) and the
Hamiltonian and  (\ref{wit4354}), the following relations:
\begin{equation}\label{relationsforsusy}
\{{\mathcal{Q}},{\mathcal{Q}}^{\dag}\}=H{\,}{\,},{\mathcal{Q}}^2=0,{\,}{\,}{{\mathcal{Q}}^{\dag}}^2=0,{\,}{\,}\{{\mathcal{Q}},W\}=0,{\,}{\,}W^2=I,{\,}{\,}[W,H]=0
\end{equation}
But the most interesting feature of this color superconductor
related supersymmetric quantum system is that the underlying $N=2$
supersymmetric quantum mechanical system, has unbroken
supersymmetry. Supersymmetry is unbroken for a quantum mechanical
system if there exists at least one quantum state in the Hilbert
space, $|\psi_{0}\rangle$,  with vanishing energy eigenvalue, that
is $H|\psi_{0}\rangle =0$. In turn, this entails that
${\mathcal{Q}}|\psi_{0}\rangle =0$ and ${\mathcal{Q}}^{\dag}|\psi_{0}\rangle =0$. For a
negative parity state this implies,
\begin{equation}\label{phi5}
|\psi^{-}_0\rangle =\left(%
\begin{array}{c}
  0 \\
  |\phi^{-}_{0}\rangle \\
\end{array}%
\right)
\end{equation}
or equivalently $A|\phi^{-}_{0}\rangle =0$. Moreover for a
positive parity ground state we have,
\begin{equation}\label{phi6s6}
|\psi^{+}_{0}\rangle =\left(%
\begin{array}{c}
  |\phi^{+}_0\rangle \\
  0 \\
\end{array}%
\right)
\end{equation}
or equivalently $A^{\dag}|\phi^{+}_{0}\rangle =0$. Whether
supersymmetry is unbroken or not, is very closely related to the
number of zero modes of the system. Zero modes are perfectly
described by the Witten index. Let $n_{\pm}$ be the number of zero
modes of $H_{\pm}$ in the subspace $\mathcal{H}^{\pm}$. For a
finite number of zero modes, $n_{+}$ and $n_{-}$, we define the
Witten index of the system to be,
\begin{equation}\label{phil}
\Delta =n_{-}-n_{+}
\end{equation}
In the case the Witten index is an non-zero integer, supersymmetry
is unbroken for sure. The case for which the Witten index is zero
is much more involved. Indeed, if the Witten index is zero, it and
if $n_{+}=n_{-}=0$ supersymmetry is broken. Conversely, if $n_{+}=
n_{-}\neq 0$ the system retains an unbroken supersymmetry. The
definition for the Witten index we just gave, holds true for
Fredholm operators only. An operator $A$ is Fredholm, if it has
discrete spectrum, a fact that is ensured if
$\mathrm{dim{\,}ker}A<\infty$. By the same reasoning, if an
operator is trace-class, this embodies the Fredholm feature for
this operator \cite{thaller}. Accordingly, the Fredholm index of
the operator $A$ is closely related to the Witten index with the
former defined as,
\begin{equation}\label{ker}
\mathrm{ind} A = \mathrm{dim}{\,}\mathrm{ker}
A-\mathrm{dim}{\,}\mathrm{ker} A^{\dag}=
\mathrm{dim}{\,}\mathrm{ker}A^{\dag}A-\mathrm{dim}{\,}\mathrm{ker}AA^{\dag}
\end{equation}
Indeed the relation between the aforementioned two indices is,
\begin{equation}\label{ker1}
\Delta=-\mathrm{ind} A=\mathrm{dim}{\,}\mathrm{ker}
H_{-}-\mathrm{dim}{\,}\mathrm{ker} H_{+}
\end{equation}
As we shall see shortly, the operators $\mathcal{D}$ and $\mathcal{D}^{\dag}$ defined
in relations (\ref{bdgeqinmform}) and (\ref{bdgeqinmformadj}) are
Fredholm for the localized solutions (the localization entails
specific boundary conditions for the operators which in the end
render the operators to be Fredholm) around the vortex. The vector
space $\mathrm{ker}\mathcal{D}$ is given by the solutions of the equation
$\mathcal{D}\Phi=0$, with the solutions $\Phi$ being zero at spatial
infinity. The last property is equivalent to searching for
localized solutions around the vortex. As we have seen earlier,
the solutions of the equation $\mathcal{D}\Phi =0$, are given by the
solutions of the equation (\ref{zerobdg}), which are the two
solutions we found earlier, namely, $\phi_R$ and $\phi_L$ and are
explicitly given by equations (\ref{rsymetry}) and
(\ref{rsymetry1}). Hence the two localized solutions constitute
the space $\mathrm{ker}\mathcal{D}$ for the operator $\mathcal{D}$. In the same line
of reasoning, the localized solutions of the equation
$\mathcal{D}^{\dag}\Phi =0$ are given by:
\begin{equation}\label{r2}
\varphi_R'=\frac{e^{i\frac{\pi}{4}}}{\sqrt{\lambda}}\left
(\begin{array}{c}
   e^{i\theta}J_1(\mu r)\\
    i J_0(\mu r)\\
    0\\
    0 \\
    J_0(\mu r) \\
    -ie^{-i\theta}J_1(\mu r) \\
   0\\
    0 \\
\end{array}\right )e^{-\int_0^r|\Delta (r')|\mathrm{d}r'}
\end{equation}
in reference to the right-handed solution, while for the left
handed one we have:
\begin{equation}\label{r3}
\varphi_L'=\frac{e^{-i\frac{\pi}{4}}}{\sqrt{\lambda}}\left
(\begin{array}{c}
    0\\
    0 \\
   e^{i\theta}J_1(\mu r)\\
    -i J_0(\mu r)\\
    0\\
    0 \\
    J_0(\mu r) \\
    i e^{-i\theta}J_1(\mu r)\\

\end{array}\right )e^{-\int_0^r|\Delta (r')|\mathrm{d}r'}
\end{equation}
In the same line of argument as in the $\mathcal{D}$ operator case, the
operator $\mathcal{D}^{\dag}$ is also Fredholm with the two solutions
$\varphi_L'$ and $\varphi_R'$ constituting the space
$\mathrm{ker}\mathcal{D}^{\dag}$. To make contact with the $N=2$ SUSY QM
algebra, the supercharges are defined in terms of the operators $\mathcal{D}$ and $\mathcal{D}^{\dag}$ and the
corresponding zero modes are classified according to their
$P^{\pm}$ parity as follows: The parity odd zero modes are (that
is the zero modes of the operator $\mathcal{D}$),
\begin{equation}\label{parodd}
|\phi^{-}_0\rangle_1=\phi_L{\,}{\,}{\,}{\,}|\phi^{-}_0\rangle_2=\phi_R
\end{equation}
while the parity even states are (the zero modes of $\mathcal{D}^{\dag}$):
\begin{equation}\label{parev}
|\phi^{+}_0\rangle_1=\varphi_L'{\,}{\,}{\,}{\,}|\phi^{+}_0\rangle_2=\varphi_R'
\end{equation}
Correspondingly, the zero modes of the Hamiltonian, $H$ are
$|\psi^{+}_0\rangle_1$, $|\psi^{+}_0\rangle_2$,
$|\psi^{-}_0\rangle_1$, $|\psi^{-}_0\rangle_2$. Since the two
operators $\mathcal{D}$ and $\mathcal{D}^{\dag}$ are Fredholm owing to the finiteness
of their corresponding $\mathrm{kernels}$, the Fredholm index of the
operator $\mathcal{D}$ is given by:
\begin{equation}\label{kerdf}
\mathrm{ind} \mathcal{D} = \mathrm{dim}{\,}\mathrm{ker}
\mathcal{D}-\mathrm{dim}{\,}\mathrm{ker} \mathcal{D}^{\dag}=
\mathrm{dim}{\,}\mathrm{ker}\mathcal{D}^{\dag}\mathcal{D}-\mathrm{dim}{\,}\mathrm{ker}\mathcal{D}\mathcal{D}^{\dag}
\end{equation}
Hence, the Witten index of the corresponding SUSY QM algebra is
given by:
\begin{equation}\label{ker1}
\Delta=-\mathrm{ind} \mathcal{D}
\end{equation}
Based on the fact that $\mathrm{ker}\mathcal{D}=\mathrm{ker}\mathcal{D}^{\dag}$ as we
found previously, the Witten index of the SUSY QM algebra is zero.
Note however that $n_{-}=n_{+}\neq 0$ (using the previously
deployed notation) a fact that implies unbroken supersymmetry (for physical systems exhibiting similar behavior, that is unbroken SUSY with zero Witten index and other interesting attributes, consult references \cite{pluskai1,pluskai2,pluskai3,pluskai4,pluskai5,pluskai6}). Let us recapitulate what we found up to now.
From the fermionic system around a vortex that is constructed by
the zero modes solutions of the Bogoliubov-de Genne equation, we
can form an $N=2$ supersymmetric quantum mechanics algebra with no
central charge. The supercharges are constructed by the operators
$\mathcal{D}$ and $\mathcal{D}^{\dag}$ which as we proved are Fredholm, in the case
the zero mode solutions are localized around the vortex.

\subsection{A Brief Comment}

Before closing this section, we will address the problem of finding the Witten index in the case we change the pairing gap $\Delta(x)$. For example let the new pairing gap $\Delta(x)'$ be related to the old pairing gap by:
\begin{equation}\label{newpair}
\Delta(x)'=\Delta(x)+\Delta_1(x)
\end{equation}
with $\Delta_1(x)=e^{i\theta}|\Delta_1(r)|$ and $\lim_{r\rightarrow \infty} |\Delta_1(r)|> 0$. At the beginning of this section we mentioned that the localized fermionic solutions around the vortex have some sort of universality, stemming from the fact that the pairing gap does not depend on $z$. This issue, has its impact on the Witten index, and in fact we shall prove that if we change the pairing gap according to relation (\ref{newpair}), the Witten index remains invariant. Hence although the solutions might change, supersymmetry remains unbroken. To see this, we shall make use of a theorem which states that, the Fredholm index of a Fredholm operator $\mathcal{D}$, namely $\mathrm{ind}\mathcal{D}$ remains invariant if we add a symmetric odd operator $C$ to this Fredholm operator, that is:
\begin{equation}\label{infgrd}
\mathrm{ind}(\mathcal{D}+C)=\mathrm{ind}\mathcal{D}
\end{equation}
In our case, since the new pairing gap obeys $\lim_{r\rightarrow \infty} |\Delta_1(r)|> 0$, the odd symmetric operator has the following representation:
\begin{equation}\label{susyqmrnmassive}
C=\left(%
\begin{array}{cc}
 0 & \Delta_1(x)
 \\ \Delta_1(x)^* & 0\\
\end{array}%
\right)
\end{equation}
Hence, the Fredholm index of the operator $\mathcal{D}$, defined in relation (\ref{bdgeqinmform}), is invariant with $\mathrm{ind}(\mathcal{D}+C)=\mathrm{ind}\mathcal{D}$. Thereby, the Witten index $\Delta=-\mathrm{ind}\mathcal{D}$  is also invariant, and hence the same results as in the case corresponding to $\Delta(x)$ hold, that is, supersymmetry is unbroken.

%\subsection{Consequences of $N=2$ Supersymmetry for the Massive Zero Modes Solutions}

%The appearance of an underlying supersymmetry for the zero modes of the fermionic system under study, has some implications for the non-zero modes of the system. By non-zero modes we mean the solutions with $E\neq 0$

%Let us see the implications
%of supersymmetry on the eigenfunctions of the Hamiltonian with
%$E\neq 0$. The Hamiltonians $H_+$ and $H_-$, are known to be
%isospectral for eigenvalues different from zero
%\cite{thaller,susyqm}, that is,
%\begin{equation}\label{isosp}
%\mathrm{spec}(H_{+})\setminus \{0\}=\mathrm{spec}(H_{-})\setminus
%\{0\}
%\end{equation}
%In addition, the following relations hold,
%\begin{equation}\label{positivee}
%Q|\psi^{-}_0\rangle=\sqrt{E}|\psi^{+}_0\rangle{\,}{\,}{\,}\mathrm{and}{\,}{\,}{\,}Q^{\dag}|\psi^{+}_0\rangle=\sqrt{E}|\psi^{-}_0\rangle,
%\end{equation}
%with $E$ the common eigenvalues of the Hamiltonians $H_+$ and
%$H_-$. In turn these imply,
%\begin{equation}\label{pose}
%D|\phi^{-}\rangle=\sqrt{E}|\phi^{+}\rangle
%{\,}{\,}{\,}\mathrm{and}{\,}{\,}{\,}D^{\dag}|\phi^{+}\rangle=\sqrt{E}|\phi^{-}\rangle.
%\end{equation}

\section{$N=2$ SUSY QM and Massless Dirac Fermion Quasinormal Modes in Reissner-Nordstr\"{o}m-anti-de Sitter black hole spacetimes}

In this section we shall present a system of Dirac fermions in a
gravitational background, from which we can construct an $N=2$
SUSY QM algebra. The gravitational background is that of an
Reissner-Nordstr\"{o}m-anti-de Sitter black hole spacetime. This
background is a potential candidate spacetime that can describe
color superconductivity. In view of the AdS/CFT correspondences
between gauge theory and gravity, the fact that the
aforementioned gravitational system and the color superconductor
fermionic system have an underlying $N=2$ SUSY QM is rather
useful. Hence, although the two models are independent at first
sight, they have a common underlying symmetry pattern which can be
useful.

\noindent To be more specific, the supersymmetry we shall present
shortly, is very closely related to the quasinormal modes of the
Dirac fermionic field in the Reissner-Nordstr\"{o}m-anti-de Sitter
black hole background. The perturbation of a black hole can be
achieved either by directly perturbing the gravitational
background or by simply adding matter or gauge fields in the black
hole spacetime \cite{roman}. In the linear approximation, the
fermionic field has no back-reaction on the metric. The metric in
a d-dimensional Reissner-Nordstr\"{o}m-anti-de Sitter spacetime is
given by:
\begin{equation}\label{RNmetric}
\mathrm{d}s^2=-f(r)\mathrm{d}t^2+\frac{1}{f(r)}\mathrm{d}r^2+r^2\mathrm{d}\Omega^2_{d-2,k}
\end{equation}
where, $f(r)$ is equal to:
\begin{equation}\label{fr}
f(r)=k+\frac{r^2}{L^2}+\frac{Q^2}{4r^{2d-6}}-\Big{(}\frac{r_0}{r}\Big{)}^{d-3}
\end{equation}
In the above equation, $L$ is the AdS radius, Q is the black hole
charge, and $r_0$ is related to the black hole mass $M$. The
$\mathrm{d}\Omega^2_{d-2,k}$ is the metric of constant curvature,
with $k$ characterizing the curvature. The value $k>0$
characterizes the metric of an $d-2$ dimensional sphere, while the
$k=0$ describes $R^{d-2}$. Finally when $k<0$ it describes
$H^{d-2}$. We shall focus on the flat case in this paper, since we
would like to make contact to a superconductor on a plane. In the
4-dimensional case, the zero curvature
Reissner-Nordstr\"{o}m-anti-de Sitter metric is,
\begin{equation}\label{metric4}
\mathrm{d}s^2=-f(r)\mathrm{t}^2+\frac{1}{f(r)}\mathrm{d}r^2+r^2(\mathrm{d}x^2+\mathrm{d}y^2)
\end{equation}
The corresponding spin connection $\omega_{\hat{a}\hat{b}c}$, is
equal to:
\begin{equation}\label{sup1}
\omega_{\hat{a}\hat{b}c}=e_{\hat{a}d}\partial_{c}e^{d}_{\hat{b}}+e_{\hat{a}d}e^{f}_{\hat{b}}\Gamma^{d}_{{\,}{\,}fc}
\end{equation}
where, $e_{\hat{a}d}$ denotes the tetrad field, while
$\Gamma^{d}_{{\,}{\,}fc}$ denotes the Christoffel connection. The
Einstein-Maxwell action for the Dirac fermion field equals to \cite{cai}:
\begin{align}\label{actionrn}
&
\mathcal{S}=\frac{1}{2G_4^2}\int\mathrm{d}^2x\sqrt{-g}\Big{(}\mathcal{R}-\frac{6}{L^2}\Big{)}
\\ \notag & +\mathcal{N}\int\mathrm{d}^4x\sqrt{-g}\Big{(}-\frac{1}{4}F_{ab}F^{ab}+i(\bar{\Psi}\Gamma^{\alpha}(D_a-iqA_a)\Psi-m\bar{\Psi}\Psi)\Big{)}
\end{align}
In the above action (\ref{actionrn}), $G_4$ is the 4-dimensional
gravitational constant, $\mathcal{R}$ is the corresponding Ricci
scalar, $\mathcal{N}$ is a total coefficient characterizing matter
fields, and $q$ is the coupling constant between the fermion field
and the abelian gauge field $A_a$. Additionally, the operator
$D_a$ is:
\begin{equation}\label{sup2}
D_a=\partial_a+\frac{1}{2}\omega_{\hat{c}\hat{b}a}\Sigma^{\hat{c}\hat{b}}
\end{equation}
with
$\Sigma^{\hat{c}\hat{b}}=\frac{1}{4}[\Gamma^{\hat{c}},\Gamma^{\hat{b}}]$,
and the Dirac gamma matrices are related to the vierbeins as, $\Gamma^b=e^b_{\hat{a}}\Gamma^{\hat{a}}$. A solution of
the equations of motion corresponding to the action
(\ref{actionrn}) is:
\begin{equation}\label{solutionrn}
A_t=Q(\frac{1}{r}-\frac{1}{r_+}),{\,}{\,}{\,}{\,}{\,}{\,}\Psi=0
\end{equation}
In order to extract the quasinormal mode spectrum corresponding to
the Reissner-Nordstr\"{o}m-anti-de Sitter black hole spacetime, we
consider the limit in which the fermionic field does not backreact
on the metric and the abelian field, as we also mentioned at the
beginning of this section. The wave function solution
$\Psi(r,x_{\mu})$ can be written in the following form \cite{cai}:
\begin{equation}\label{functionpsi}
\Psi(r,x_{\mu})=\psi(r)e^{-i\omega t+i\vec{k}{\,}{\,}\cdot
\vec{x}}
\end{equation}
with $x_{\mu}=(t,x,y)$ and $\vec{x}=(x,y)$. Using the above form
of the function (\ref{functionpsi}), the Dirac equation can be
cast into the following form \cite{cai}:
\begin{equation}\label{dequatafter}
\sqrt{f}\Gamma^{\hat{r}}\partial_r \psi-\frac{i
\omega}{\sqrt{f}}\Gamma^{\hat{t}}\psi+\frac{i\vec{k}\cdot
\Gamma^{\hat{\vec{x}}}}{r}\psi+\frac{1}{4}\Big{(}\frac{f'}{\sqrt{f}}+\frac{4\sqrt{f}}{r}\Big{)}\Gamma^{\hat{r}}\psi-(iq\Gamma^{\alpha}A_{\alpha}+m)\psi=0
\end{equation}
where $\vec{k}\cdot
\Gamma^{\hat{\vec{x}}}=k_x\Gamma^{\hat{x}}+k_y\Gamma^{\hat{\vec{y}}}$.
The Dirac gamma matrices can be written in the following
representation:
\begin{equation}\label{dgammama}
\Gamma^{\hat{t}}=\bigg{(}\begin{array}{ccc}
  I & 0 \\
  0 & -I  \\
\end{array}\bigg{)},{\,}{\,}{\,}{\,}\Gamma^{\hat{i}}=\bigg{(}\begin{array}{ccc}
  0 & \sigma^{\hat{i}} \\
  \sigma^{\hat{i}} & 0  \\
\end{array}\bigg{)}
\end{equation}
with $I$ the identity matrix and $\sigma^{i}$ the Pauli matrices,
namely:
\begin{equation}\label{sigmapauli}
\sigma^{\hat{x}}=\bigg{(}\begin{array}{ccc}
  0 & 1 \\
  1 & 0  \\
\end{array}\bigg{)},{\,}{\,}{\,}{\,}\sigma^{\hat{y}}=\bigg{(}\begin{array}{ccc}
  0 & -i \\
  i & 0  \\
\end{array}\bigg{)},{\,}{\,}\sigma^{\hat{r}}=\bigg{(}\begin{array}{ccc}
  1 & 0 \\
  0 & -1 \\
\end{array}\bigg{)}
\end{equation}
For later convenience, we decompose the fermion field Hilbert
space to the chirality operator subspaces, that is:
\begin{equation}\label{chiral1}
\Psi=\bigg{(}\begin{array}{ccc}
  \Psi_{+} \\
  \Psi_{-}  \\
\end{array}\bigg{)}
\end{equation}
and $P_{\pm}\Psi=\pm\Psi_{\pm}$, with $P_{\pm}=1\pm \Gamma^5$ and
$\Gamma^5=i\Gamma^t\Gamma^x\Gamma^y\Gamma^r$. Using the eigenstates
$\Psi_{\pm}$, the Dirac equations of motion can be cast as \cite{cai}:
\begin{equation}\label{deqm2}
\sqrt{f}\partial_r+\frac{1}{4}\frac{f'}{\sqrt{f}}+\frac{\sqrt{f}}{r}\sigma^{\hat{r}}\psi_{-}\frac{i}{r}(\vec{k}\cdot
\vec{\sigma})\psi_{-}+\frac{i}{\sqrt{f}}(\omega+qA_t)\psi_{-}-m\psi_+=0
\end{equation}
 and also
\begin{equation}\label{deqm22}
\sqrt{f}\partial_r+\frac{1}{4}\frac{f'}{\sqrt{f}}+\frac{\sqrt{f}}{r}\sigma^{\hat{r}}\psi_{+}\frac{i}{r}(\vec{k}\cdot
\vec{\sigma})\psi_{+}+\frac{i}{\sqrt{f}}(\omega+qA_t)\psi_{-}-m\psi_{-}=0
\end{equation}
with $\Psi_+=\psi_+e^{-i\omega t+i\vec{k}\vec{x}}$ and
$\Psi_-=\psi_{-}e^{-i\omega t+i\vec{k}\vec{x}}$. The set of the
above equations (\ref{deqm2}) and (\ref{deqm22}) is invariant
under the transformation:
\begin{equation}\label{transfo}
\omega \rightarrow -\omega,{\,}{\,}{\,}{\,}{\,}q\rightarrow -q,
{\,}{\,}{\,}{\,}{\,}\psi_{+}\rightarrow \psi_{-}
\end{equation}
In the rest of this paper we shall be interested in the chiral
limit $m=0$. This will result to an unbroken chiral symmetry for
the system, which proves to be very important and could be an
underlying link between the fermionic gravitational system and the
color superconductor around a vortex system. We focus on the
quasinormal modes of $\psi_{+}$. We set $k_y=0$. This is because
the symmetry that the system possesses on the
$(\vec{x},\vec{y})$-plane. Upon rewriting $\psi_{+}$ as
$\psi_{+}=r^{-1}f^{-1/4}\tilde{\psi}$, the equation (\ref{deqm22})
can be simplified to the following one:
\begin{equation}\label{simeqd}
\sigma^{\hat{r}}\tilde{\psi}'-\frac{i}{f}(\omega+qA_t-\frac{\sqrt{f}}{r}k_{x}\sigma^{\hat{x}})\tilde{\psi}=0
\end{equation}
Decomposing the fermionic field $\tilde{\psi}$, as:
\begin{equation}\label{decomp}
\tilde{\psi}=\bigg{(}\begin{array}{ccc}
  \psi_{1} \\
  \psi_{2}  \\
\end{array}\bigg{)}
\end{equation}
the above equation (\ref{simeqd}), can be recast in the following
form:
\begin{align}\label{decver222}
&
\psi_1'-\frac{i}{f}(\omega+qA_t)\psi_1+\frac{i}{r\sqrt{f}}k_x\psi_2=0
\\ \notag &
\psi_2'+\frac{i}{f}(\omega+qA_t)\psi_2-\frac{i}{r\sqrt{f}}k_{x}\psi_1=0
\end{align}
The above equations are invariant under the symmetry:
\begin{equation}\label{transfo2}
\omega \rightarrow -\omega,{\,}{\,}{\,}{\,}{\,}q\rightarrow -q,
{\,}{\,}{\,}{\,}{\,}k_{x}\rightarrow -k_{x},
{\,}{\,}{\,}{\,}{\,}\psi_{1}\rightarrow \psi_{2}
\end{equation}
Using these equations, namely (\ref{decver222}) we can construct an
$N=2$ supersymmetric algebra. This algebra is founded on the
matrix:
\begin{equation}\label{susyqmrn567m12}
{{{{{\mathcal{D}}_{RN}}}}}=\left(%
\begin{array}{cc}
 \partial_r-\frac{i}{f}(\omega+qA_t) & \frac{i}{r\sqrt{f}}k_x
 \\  -\frac{i}{r\sqrt{f}}k_x & \partial_r+\frac{i}{f}(\omega+qA_t)\\
\end{array}%
\right)
\end{equation}
acting on the vector:
\begin{equation}\label{ait34e1}
\left(%
\begin{array}{c}
 \psi_1 \\
  \psi_2 \\
\end{array}%
\right)
\end{equation}
It is obvious that the zero modes of the matrix
(\ref{susyqmrn567m12}) yield the solutions of equation
(\ref{decver222}) with respect to $\omega$. But these solutions
correspond to the zero modes of the Dirac fermionic system.
Therefore the zero mode solutions of the matrix
(\ref{susyqmrn567m12}) and the quasinormal modes of the Dirac
fermionic system are in bijective correspondence. Thereby, the
existence of quasinormal modes guarantees the existence of zero
modes for the aforementioned matrix. The adjoint of the matrix
${{{{{\mathcal{D}}_{RN}}}}}$ is equal to:
\begin{equation}\label{adjsusyqmrn567m}
{{{{{\mathcal{D}}_{RN}}}}}^{\dag}=\left(%
\begin{array}{cc}
 \partial_r+\frac{i}{f}(\omega^*+qA_t) & \frac{i}{r\sqrt{f}}k_x
 \\  -\frac{i}{r\sqrt{f}}k_x & \partial_r-\frac{i}{f}(\omega^*+qA_t)\\
\end{array}%
\right)
\end{equation}
Correspondingly, the supercharges of the $N=2$ algebra ${\mathcal{Q}}_{RN}$
and $Q^{\dag}_{RN}$  are equal to:
\begin{equation}\label{wit2dr}
{\mathcal{Q}}_{RN}=\bigg{(}\begin{array}{ccc}
  0 & {{{{{\mathcal{D}}_{RN}}}}} \\
  0 & 0  \\
\end{array}\bigg{)}, {\,}{\,}{\,}{\,}{\,}Q^{\dag}_{RN}=\bigg{(}\begin{array}{ccc}
  0 & 0 \\
  {{{{{\mathcal{D}}_{RN}}}}}^{\dag} & 0  \\
\end{array}\bigg{)}
\end{equation}
Moreover, the quantum Hamiltonian is equal to,
\begin{equation}\label{wit4354dr}
H_{RN}=\bigg{(}\begin{array}{ccc}
  {{{{\mathcal{D}}_{RN}}}}{{{{\mathcal{D}}_{RN}}}}^{\dag} & 0 \\
  0 & {{{{\mathcal{D}}_{RN}}}}^{\dag}{{{{\mathcal{D}}_{RN}}}}  \\
\end{array}\bigg{)}
\end{equation}
The supercharges (\ref{wit2dr}) and the Hamiltonian and
(\ref{wit4354dr}), satisfy the equations (\ref{s23}) and
(\ref{s45}), namely
\begin{equation}\label{structureqns}
\{{\mathcal{Q}}_{RN},{\mathcal{Q}}_{RN}^{\dag}\}=H_{RN},{\,}{\,}{\mathcal{Q}}_{RN}^2=0,{\,}{\,}{{\mathcal{Q}}_{RN}^{\dag}}^2=0,{\,}{\,}\{{\mathcal{Q}}_{RN},W\}=0,{\,}{\,}W^2=I,{\,}{\,}[W,H_{RN}]=0
\end{equation}
 Hence the algebraic structure of an $N=2$ SUSY QM algebra, underlies this fermionic system that corresponds to the solution $\psi_+$ (recall that there is another identical system corresponding to $\psi_{-}$, which we describe soon). Let's see if this underlying supersymmetry is broken or unbroken. The last strongly depends on the index of the operator ${{{{\mathcal{D}}_{RN}}}}$. But since the number of quasinormal modes is a discrete infinite set, and owing to the bijective correspondence between the zero modes of the operator ${\mathcal{D}}_{RN}$ and the quasinormal modes, we conclude that the zero modes form a discrete infinite set. Therefore, the operator ${\mathcal{D}}_{RN}$ is not Fredholm which means that the index of the operator and correspondingly the Witten index must be regularized.
In order to do so, we shall make use of the heat-kernel
regularized index \cite{witten,thaller,susyqm,susyqm1}, both for the operator ${\mathcal{D}}_{RN}$, denoted
$\mathrm{ind}_t{\mathcal{D}}_{RN}$ and for the Witten index, $\Delta_t$, which
are defined  as:
\begin{align}\label{heatkerw}
&
\mathrm{ind}_t{\mathcal{D}}_{RN}=\mathrm{Tr}(-We^{-t{\mathcal{D}}_{RN}^{\dag}{\mathcal{D}}_{RN}})=\mathrm{tr}_{-}(-We^{-t{\mathcal{D}}_{RN}^{\dag}{\mathcal{D}}_{RN}})-\mathrm{tr}_{+}(-We^{-t{\mathcal{D}}_{RN}{\mathcal{D}}_{RN}^{\dag}})
\\ \notag & \Delta_t=\lim_{t\rightarrow
\infty}\mathrm{ind}_t{\mathcal{D}}_{RN}
\end{align}
The parameter $t$, is positive number $t>0$, and  moreover the
trace $\mathrm{tr}_{\pm }$, stands for the trace in the subspaces
$\mathcal{H}^{\pm}$. The heat-kernel regularized index is defined
for   trace class operators \cite{thaller}. In the regularized
index case, the same hold in reference to supersymmetry breaking,
that is if $\mathrm{\Delta}_t\neq 0$ supersymmetry is unbroken. When
the Witten index is zero, if
$\mathrm{ker}{\mathcal{D}}_{RN}=\mathrm{ker}{\mathcal{D}}_{RN}^{\dag}=0$, supersymmetry is
broken, while when
$\mathrm{ker}{\mathcal{D}}_{RN}=\mathrm{ker}{\mathcal{D}}_{RN}^{\dag}\neq0$ supersymmetry
is unbroken. In the case at hand, supersymmetry is unbroken. We
can see this without solving the zero mode equation of the
${\mathcal{D}}_{RN}^{\dag}$ operator. Indeed the existence of zero modes
suffices to argue about supersymmetry. Since
$\mathrm{ker}{\mathcal{D}}_{RN}\neq 0$,  the zero modes equation for the
operator ${\mathcal{D}}_{RN}^{\dag}$ can yield two results. Either that
$\mathrm{ker}{\mathcal{D}}_{RN}^{\dag}\neq 0$ or that
$\mathrm{ker}{\mathcal{D}}_{RN}^{\dag}=0$. If the second is true, then the
Witten index is different than zero, $\Delta_t\neq 0$, hence
supersymmetry is unbroken. In the first case,
$\mathrm{ker}{\mathcal{D}}_{RN}^{\dag}\neq 0$, it can either be that
$\mathrm{ker}{\mathcal{D}}_{RN}^{\dag}=\mathrm{ker}{\mathcal{D}}_{RN}$ or that
$\mathrm{ker}{\mathcal{D}}_{RN}^{\dag}\neq \mathrm{ker}{\mathcal{D}}_{RN}$. In both cases
supersymmetry is unbroken. Hence the system that is described by
the $\psi_{+}$ function has an  underlying unbroken $N=2$
supersymmetry.

Recall that there is another solution to the Dirac equation in
this curved background, namely $\psi_{-}$. The equations of motion
corresponding to $\psi_{-}$ are equal to:
\begin{align}\label{decver2}
&
\psi_1'-\frac{i}{f}(-\omega-qA_t)\psi_1'-\frac{i}{r\sqrt{f}}k_x\psi'_2=0
\\ \notag &
\psi_2'+\frac{i}{f}(-\omega-qA_t)\psi_2'+\frac{i}{r\sqrt{f}}k_{x}\psi_1'=0
\end{align}
with,
\begin{equation}\label{decomp1}
\tilde{\psi}'=\bigg{(}\begin{array}{ccc}
  \psi_{1}' \\
  \psi_{2}'  \\
\end{array}\bigg{)}
\end{equation}
and $\psi_{-}$ being related to
$\psi_{-}=r^{-1}f^{-1/4}\tilde{\psi}'$. By the same reasoning as
in the $\psi_{+}$ case, the supersymmetric quantum algebra can be
built on the matrix:
\begin{equation}\label{susyqmrn567m}
{\mathcal{D}}_{RN'}=\left(%
\begin{array}{cc}
 \partial_r-\frac{i}{f}(-\omega-qA_t) & -\frac{i}{r\sqrt{f}}k_x
 \\  \frac{i}{r\sqrt{f}}k_x & \partial_r+\frac{i}{f}(-\omega-qA_t)\\
\end{array}%
\right)
\end{equation}
acting on the vector:
\begin{equation}\label{ait34e1}
\left(%
\begin{array}{c}
 \psi_1' \\
  \psi_2' \\
\end{array}%
\right)
\end{equation}
The supercharges of the new algebra are equal to
\begin{equation}\label{wit2dr1}
{\mathcal{Q}}_{RN'}=\bigg{(}\begin{array}{ccc}
  0 & {\mathcal{D}}_{RN'} \\
  0 & 0  \\
\end{array}\bigg{)}, {\,}{\,}{\,}{\,}{\,}Q^{\dag}_{RN'}=\bigg{(}\begin{array}{ccc}
  0 & 0 \\
  {\mathcal{D}}_{RN'}^{\dag} & 0  \\
\end{array}\bigg{)}
\end{equation}
and the Hamiltonian is
\begin{equation}\label{wit4354dr1231}
H_{RN'}=\bigg{(}\begin{array}{ccc}
  {\mathcal{D}}_{RN'}{\mathcal{D}}_{RN'}^{\dag} & 0 \\
  0 & {\mathcal{D}}_{RN'}^{\dag}{\mathcal{D}}_{RN'}  \\
\end{array}\bigg{)}
\end{equation}
Following the same line of argument as in the previous, we easily
find an $N=2$ underlying supersymmetry. Denoting the algebra
corresponding to $\psi_{-}$, $N_2$ and the one corresponding to
$\psi_1$, $N_1$ we have come to the result that the Dirac
fermionic system in an Reissner-Nordstr\"{o}m-anti-de Sitter
background, possesses an supersymmetry $N$, that is the direct sum
of two $N=2$ supersymmetries, namely:
\begin{equation}\label{directsum}
N_{total}=N_1\oplus N_2
\end{equation}
The total Hamiltonian of the system is $H_{total}=H_{RN'}+H_{RN}$.
It is tempting to investigate if this supersymmetry $N_{total}$
results after the breaking of a larger supersymmetry, for example
an $N=4$ supersymmetry, or even the possibility that a central
charge exists. In addition this $N_{total}$ supersymmetry could be
the a sign of an underlying higher symmetry (for a quite similar situation but in a different context, consult reference \cite{ivanovlan1} where two $N=2$, $d=1$ supersymmetries constitute a $N=4$ supersymmetry). We defer this investigation to a future publication. Let us just
mention that the $N=4$ supersymmetric algebra is very important in
string theory, since extended (with $N=4,6...$) supersymmetric
quantum mechanics models are the resulting models from the
dimensional reduction to one (temporal) dimension of $N=2$ and
$N=1$ Super-Yang Mills theories
\cite{ivanov1,ivanov2,ivanov3,ivanov4,ivanov5,ivanov6,ivanov7,ivanov8,ivanov9}.
In addition, extended supersymmetries serve as superextensions of
integrable models like Calogero-Moser systems, Landau-type models
\cite{ivanovlan} and also there exist interesting dualities
between various supermultiplets with string theory origin (like
T-duality) \cite{spector}. But the most salient feature of the extended supersymmetric quantum algebra is that it can be
connected to a generalized harmonic superspace
\cite{ivanov1,ivanov2,ivanov3,ivanov4,ivanov5,ivanov6,ivanov7,ivanov8,ivanov9},
with the last being a powerful tool for $N\geq 4$ supersymmetric
model building. These harmonic space structures are linked to
supersymmetric linear models defined in the target space, for
which the harmonic variables give rise to target space harmonics.
Note that the Dirac solutions, actually the zero modes of the
supercharges, are sections of the total spin bundle \cite{nakahara,jost} over the
Riemannian manifold $M$. Hence we can directly connect these sections to an
extended supersymmetric sigma model in harmonic superspace. Although
these issues are very interesting both in mathematical and
physical aspects, we defer this work to a future article where we
address formally all the aforementioned topics.

\noindent The two $N=2$ supersymmetries can be combined to a higher representation of a single $N=2$, $d=1$ supersymmetry. Indeed, the supercharges of this representation, which we denote ${\mathcal{Q}}_{T}$ and  ${\mathcal{Q}}_{T}^{\dag}$ is equal to:
\begin{equation}\label{connectirtyrtons}
{\mathcal{Q}}_{T}= \left ( \begin{array}{cccc}
  0 & 0 & 0 & 0 \\
  {\mathcal{D}}_{RN} & 0 & 0 & 0 \\
0 & 0 & 0 & 0 \\
0 & 0 & {\mathcal{D}}_{RN'}^{\dag} & 0  \\
\end{array} \right),{\,}{\,}{\,}{\,}{\mathcal{Q}}_{T}^{\dag}= \left ( \begin{array}{cccc}
  0 &  {\mathcal{D}}_{RN}^{\dag} & 0 & 0 \\
  0 & 0 & 0 & 0 \\
0 & 0 & 0 & {\mathcal{D}}_{RN'} \\
0 & 0 & 0 & 0  \\
\end{array} \right)
\end{equation}
Accordingly, the Hamiltonian of the combined quantum system, which we denote $H_T$ reads,
\begin{equation}\label{connections1dtr}
H_T= \left ( \begin{array}{cccc}
  {\mathcal{D}}_{RN}^{\dag}{\mathcal{D}}_{RN} & 0 & 0 & 0 \\
  0 & {\mathcal{D}}_{RN}{\mathcal{D}}_{RN}^{\dag} & 0 & 0 \\
0 & 0 & {\mathcal{D}}_{RN'}{\mathcal{D}}_{RN'}^{\dag} & 0 \\
0 & 0 & 0 & {\mathcal{D}}_{RN'}^{\dag}{\mathcal{D}}_{RN'}  \\
\end{array} \right)
\end{equation}
The operators (\ref{connectirtyrtons}) and (\ref{connections1dtr}), satisfy the $N=2$, $d=1$ supersymmetric quantum mechanics algebra, namely:
\begin{equation}\label{mousikisimagne}
\{ {\mathcal{Q}}_{T},{\mathcal{Q}}_{T}^{\dag}\}=H_{T},{\,}{\,}{\mathcal{Q}}_{T}^2=0,{\,}{\,}{{\mathcal{Q}}_{T}^{\dag}}^2=0,{\,}{\,}\{{\mathcal{Q}}_{T},W\}=0,{\,}{\,}W_T^2=I,{\,}{\,}[W_T,H_{T}]=0
\end{equation}
In this case, the Witten parity operator is equal to:
\begin{equation}\label{wparityopera}
W= \left ( \begin{array}{cccc}
  1 & 0 & 0 & 0 \\
  0 & -1 & 0 & 0 \\
0 & 0 & 1 & 0 \\
0 & 0 & 0 & -1  \\
\end{array} \right)
\end{equation}
There exist equivalent higher dimensional representations for the combined $N=2$, $d=1$ algebra, which can be obtained from the above algebra, by making the following set of replacements:
\begin{equation}\label{setof transformations}
\mathrm{Set}{\,}{\,}{\,}A:{\,}
\begin{array}{c}
 {\mathcal{D}}_{RN}\rightarrow {\mathcal{D}}_{RN}^{\dag} \\
  {\mathcal{D}}_{RN'}^{\dag}\rightarrow {\mathcal{D}}_{RN'} \\
\end{array},{\,}{\,}{\,}\mathrm{Set}{\,}{\,}{\,}B:{\,}
\begin{array}{c}
 {\mathcal{D}}_{RN}\rightarrow {\mathcal{D}}_{RN'}^{\dag} \\
  {\mathcal{D}}_{RN'}^{\dag}\rightarrow {\mathcal{D}}_{RN} \\
\end{array},{\,}{\,}{\,}\mathrm{Set}{\,}{\,}{\,}C:{\,}
\begin{array}{c}
 {\mathcal{D}}_{RN}\rightarrow {\mathcal{D}}_{RN'} \\
  {\mathcal{D}}_{RN'}^{\dag}\rightarrow {\mathcal{D}}_{RN}^{\dag} \\
\end{array}
\end{equation}

\subsection{A Global $U(1)\times U(1)$}

As we saw previously, the fermionic system in the
Reissner-Nordstr\"{o}m-anti-de Sitter black hole background can
constitute a space with two $N=2$ supersymmetric quantum mechanics
algebras, described by the supercharges defined in relations
(\ref{wit2dr}) and (\ref{wit2dr1}). The two superalgebras are
invariant under the transformations:
\begin{align}\label{transformationu1}
& {\mathcal{Q}}_{RN}^{'}=e^{-ia}{\mathcal{Q}}_{RN}, {\,}{\,}{\,}{\,}{\,}{\,}{\,}{\,}
{\,}{\,}{\mathcal{Q}}^{'\dag}_{RN}=e^{ia}{\mathcal{Q}}^{\dag}_{RN} \\ \notag &
{\mathcal{Q}}_{RN'}^{'}=e^{-ia'}{\mathcal{Q}}_{RN'}, {\,}{\,}{\,}{\,}{\,}{\,}{\,}{\,}
{\,}{\,}{\mathcal{Q}}^{'\dag}_{RN'}=e^{ia'}{\mathcal{Q}}^{\dag}_{RN'}
\end{align}
Thus each quantum system is invariant under an $R$-symmetry of the
form of an global-$U(1)$. Correspondingly, the total system is
invariant under an $U(1)\times U(1)$ symmetry. Each of the aforementioned
$U(1)$ symmetries is a symmetry of the Hilbert states
corresponding to the spaces $\mathcal{H_{RN}}^{+}$,
$\mathcal{H_{RN}}^{-}$ and $\mathcal{H_{RN'}}^{+}$,
$\mathcal{H_{RN'}}^{-}$. Let, $\psi^{+}_{RN}$ and
$\psi^{-}_{RN}$ denote the Hilbert states corresponding to the
spaces $\mathcal{H}^{+}_{RN}$ and $\mathcal{H}^{-}_{RN}$. The
$U(1)$ transformation of the states is equal to,
\begin{equation}
\psi^{'+}_{RN}=e^{-i\beta_{+}}\psi^{+}_{RN},
{\,}{\,}{\,}{\,}{\,}{\,}{\,}{\,}
{\,}{\,}\psi^{'-}_{RN}=e^{-i\beta_{-}}\psi^{-}_{RN}
\end{equation}
Obviously the parameters $\beta_{+}$ and $\beta_{-}$ are global
parameters so that $a=\beta_{+}-\beta_{-}$. Accordingly, for the
spaces  $\mathcal{H_{RN'}}^{+}$, $\mathcal{H_{RN'}}^{-}$ we have,
\begin{equation}
\psi^{'+}_{RN'}=e^{-i\beta_{+}'}\psi^{+}_{RN'},
{\,}{\,}{\,}{\,}{\,}{\,}{\,}{\,}
{\,}{\,}\psi^{'-}_{RN'}=e^{-i\beta_{-}'}\psi^{-}_{RN'}
\end{equation}
with $\psi^{+}_{RN'}$ and $\psi^{-}_{RN'}$ the Hilbert states of
the spaces $\mathcal{H_{RN'}}^{+}$, $\mathcal{H_{RN'}}^{-}$
respectively. It worths mentioning that in some superconductors,
such $U(1)$ symmetries are realized. Particularly an initial
$U(1)\times U(1)$ symmetry is broken to a single $U(1)$ (see \cite{odagiri}
for details).

\section*{Conclusions}

In this paper we studied the supersymmetry structure underlying
two physical fermionic systems, namely the color superconductor in
the chiral limit, around the boundary vortex and a fermionic
system in the Reissner-Nordstr\"{o}m-anti-de Sitter black hole
spacetime. For the first system we found by analyzing the
Bogoliubov-de Genne equation that, in the chiral limit, the
localized fermion zero modes of the color superconductor
constitute an $N=2$ supersymmetric quantum mechanics algebra with
zero supercharge. Interestingly, by analyzing the quasinormal
modes of the gravitational fermionic system in the
Reissner-Nordstr\"{o}m-anti-de Sitter background, we found two
unbroken $N=2$ supersymmetries. We stressed the fact that this result is interesting from a mathematical point of view, owing to the fact that
we can relate the fermionic gravitational system to a sigma model
in harmonic superspace. Note that the unbroken supersymmetry of
the system is guaranteed by the very own existence of fermionic
quasinormal modes of the gravitational system. This, in turn
has its own intrinsic appeal since quasinormal modes depend only
on a few physical parameters of the black hole. Since these
parameters enter the quantum algebra we have a supersymmetry
depending on a few physical parameters and that depends on the
existence of quasinormal modes. Moreover, the two $N=2$, $d=1$ algebras can be combined to form a higher dimensional reducible representation of an $N=2$ supersymmetric quantum mechanics algebra.

\noindent The two fermionic systems we presented in detail are believed to
be interrelated, with the fermionic system in curved background
being a promising candidate for describing the color
superconductor. Thus, the supersymmetries we found, show us that
under certain very general assumptions, these two systems have an
underlying supersymmetric structure, of $N=2$, $d=1$ type. Hence,
this common, in some way, underlying theme makes us believe that
the two systems might be connected. But this is just an indication
and not a direct correspondence between the two models.
Additionally the supersymmetries of the two spaces have different
Hilbert space structure, a fact that is seen easily from the
operators being Fredholm in the color superconductor case, and non
Fredholm in the other case. Moreover, the gravitational system has
a much more rich structure, and both systems can be related to
extended supersymmetric algebras. In addition, it would be of particular importance to investigate whether such a supersymmetric structure exists in the case when bosonic quasinormal modes \cite{ref1,ref2,ref3,ref4,ref5,ref6} are studied. Supersymmetry for bosonic systems in curved background occurs in the context of local $N=2$ supersymmetric backgrounds, where the fluctuations of the bosonic fields evolved in an Abelian Chern Simons vortex, have the same supersymmetric quantum mechanics algebra as the fermionic system \cite{futurework}.

\end{document}